\documentclass[a4paper, amsfonts, amssymb, amsmath, reprint, showkeys,
nofootinbib, twoside]{revtex4-1}
\usepackage[T2A]{fontenc}
\usepackage[utf8]{inputenc}
\usepackage{graphicx}
\usepackage{amsmath}
\usepackage{amssymb}
\usepackage{natbib}
\usepackage{hyperref}
 \usepackage{amsfonts}

\def\apj{ApJ}
\def\apjl{ApJL}
\def\apjs{ApJ Suppl.}

\def\aap{A\&A}
\def\aaps{A\&A Suppl.}

\def\aapr{A\&ARv}

\def\mnras{MNRAS}

\def\nat{Nature}

\begin{document}


\title{Population synthesis of exoplanets accounting for orbital variations due to stellar evolution}

\author{Andriushin A.S.$^1$, Popov S.B.$^{1,2}$}
    \email[Correspondence email address: ]{andrushin@gmail.com, sergepolar@gmail.com}
    \affiliation{$^1$ Department of physics, Lomonosov Moscow State University, $^2$Sternberg Astronomical Institute} 

\begin{abstract}

In this work, the evolution of exoplanet orbits at the late stages of stellar evolution is studied by the method of population synthesis. The evolution of stars is traced from the Main Sequence stage to the white dwarf stage. The MESA package is used to calculate evolutionary tracks. The statistics of absorbed, ejected, and surviving planets by the time of the transformation of parent stars into white dwarfs are calculated taking into account the change in the rate of star formation in the Galaxy over the entire time of its existence. Planets around stars in the range of initial masses 1-8 $M_\odot$ are considered since less massive stars do not have time to leave the Main Sequence during the lifetime of the Galaxy, and more massive ones do not lead to the formation of white dwarfs. It is shown that with the initial $a$~--~$M_\mathrm{pl}$ distribution of planets adopted in this work, most (about 60\%) of the planets born from stars in the mass range under study are absorbed by their parent stars at the giant stage. A small fraction of the planets (less than one percent) are ejected from their systems because of the mass loss due to the stellar wind. The estimated number of ejected planets with masses ranging from 0.04 Earth masses to 13 Jupiter masses in the Milky way is approximately equal to 300 million.


\end{abstract}


\maketitle

\section{Introduction}
About three decades have passed since the discovery of the first exoplanets \cite{WolszczanFrail},  \cite{MayorQueloz}. During this time, the number of confirmed extrasolar planets discovered with the help of such instruments as Kepler, HARPS, HIRES, TESS, etc., exceeded 4300. Of these, more than a hundred are planets around evolved stars: red giants and subgiants. The statistics of the detection of planets around white dwarfs are more modest. We can mention a planet around the star WD 0806-661 \cite{WD0806-661}, a recently discovered candidate for WD 1856+534 \cite{WD1856+534}, as well as objects in binary systems ``white dwarf -- Main Sequence star'' (NN Ser, Gliese 86).
However, there are many more examples -- of the order of 1000 -- of ``planetary debris'' detection around white dwarfs and in their atmospheres, and those are products of planets and/or asteroid destruction. Such conclusions can be made from the analysis of the observed chemical composition of the atmospheres of dwarfs and the observations of their circumstellar disks of dust and rock fragments \cite{WD_debris}, \cite{WD_debris2}.

Thus, it can be considered an established fact that objects of planetary masses can not only remain in the system after the star has shed its envelope at the later stages of evolution but also go into low orbits around a compact object. This makes it relevant to analyze the properties of planets in the late stages of evolution and their previous history.

In order to adequately interpret the growing amount of data on exoplanets around evolved stars and to be able to judge from these data what kind of evolution the observed planetary system has undergone, it is necessary to theoretically understand the processes that determine the evolution of planetary systems, including those stages when their parent star retires from the Main sequence (MS). A model of the evolution of planetary systems under the influence of the evolution of their parent stars would make it possible to make assumptions about the past of these systems in relation to the discovered and observed planets around evolved stars. In addition, it is desirable that the model also has a predictive potential for planetary systems around MS stars.

Over the past 10 years, a lot of studies have been devoted to modeling the evolution of planetary systems of stars after the MS stage. Key results and unresolved issues are discussed, for example, in the review \cite{veras16}. The evolution of planetary systems in the late stages of a star's life occurs under the influence of various factors and at different levels, depending on the value of the semimajor axis of the orbit and the mass of a substellar object (for example, a planet or asteroid) during the life of a star on the MS, and on how this object in the future --- after the star leaves the MS, --- can be influenced by factors such as mass loss by the parent star, tidal effects in a ``star-planet'' system, radiation (Yarkovsky effect, YORP-effect), and magnetic fields.

The impact of these factors can manifest itself both in a change in the orbit of a substellar object and in a change in its physical parameters (mass and size, temperature, composition of the surface and atmosphere, etc.). The impact can be so strong that the object will be ejected from the system, or it may happen that at the giant stage, the parent star absorbs it and it ceases to exist. Thus, in this regard, it is worth mentioning that in addition to the already mentioned examples of planets around white dwarfs and giant stars, there are also examples of free-floating planets: WISE J085510.83-071442.5 \cite{WISEJ085510}, SDSS J1110+0116 \cite{SDSSJ111010.01+0116}, PSO J318.5-22 \cite{PSOJ318.5-22} and others. Also the discovery of a free-floating terrestrial planet \cite{terrestrial_rogue} deserves a special mention.
 The number of discovered free-floating planets is growing, and among them there may be those that became unbound after they were ejected from their parent planetary systems as a result of the mass loss of a star due to a strong stellar wind.

The final fate of planetary systems is determined not only by stellar evolution but also by the initial parameters of the planets. There is a large number of modern studies that are devoted to the theory of planet formation and modeling of planetary systems (see a review in \cite{Morbidelli2020}).
 Along with a detailed study of individual systems (for example, the Solar System) or the development of details of various stages of the formation of planets and the evolution of their orbits, an important place is occupied by the construction of population models, which at a coarser level include the processes of formation and evolution of objects in a wide range of initial parameters. The population synthesis of exoplanets is discussed, for example, in the works of Christophe Mordasini, Jan Alibert et al. \cite{mordasini18}, \cite{alibert13}, \cite{mordasini09}, \cite{new_mordasini}.
In our article, we actively use the results of these studies.


The aim of this work is to model planetary orbits accounting for the evolution of a star after the MS stage, as well as to calculate the statistics of absorbed, ejected from the system, and surviving planets by the time their parent star turns into a white dwarf. Calculating properties of the Galactic population of planets we take into account the history of star formation in the Milky Way.

In section \ref{model}  we present the model that underlies our population synthesis, describe the initial distributions of planets over masses and orbits carried out in our simulation, as well as the evolutionary models of stars used in the work. Then, in section \ref{code} we briefly describe the code for population synthesis written in the MatLab package. Section  \ref{results}  is devoted to the results of our study. In section  \ref{disc}  we discuss our results and approach. The final section briefly summarizes the main results of this study.



\section{Model}
\label{model}

The population synthesis and modeling of the evolution of exoplanetary systems carried out in this work are based on modern approaches to formation of planetary systems and stellar evolution, as well as on a simple model that links the evolution of a star and changes in orbits of the planets. This simple model does not take into account possible orbital variations resulting from interplanetary gravitational interactions. As for binary and multiple star systems, the model is suitable only for that fraction of them for which the distance between the parent star and the planet significantly exceeds the distance to the second star in the system (for binary systems, where a planet orbits at a large distance from a pair of stars close to each other, the model does not work since in this case the evolutionary tracks of stars due to mutual influence in many cases will be different from those used in this study).


\subsection{Model of the orbital evolution}

The problem of orbital evolution due to isotropic mass loss by a central more massive body is well known. Changes in the semimajor axis of the orbit, eccentricity, and true anomaly with time are generally described by the following differential equations (see \cite{greatescape}, \cite{1963}):


\begin{equation}
\label{dadt}
    \frac{da}{dt} = a\frac{1+e^2+2e\cos(f)}{1-e^2}\frac{\dot M_\mathrm{tot}}{M_\mathrm{tot}}  
\end{equation}
\begin{equation}
\label{dedt}
    \frac{de}{dt} = (e+ \cos(f)) \, \frac{\dot M_\mathrm{tot}}{M_\mathrm{tot}} 
\end{equation}
\begin{equation}
\label{dfdt}
    \frac{df}{dt} = - \frac{\sin(f)}{e} \frac{\dot M_\mathrm{tot}}{M_\mathrm{tot}}  +\frac{n (1+e\cos(f))^2}{(1-e^2)^{3/2}},
\end{equation} 
where $f$ -- is the true anomaly of the orbit, $a$ -- its semimajor axis,  $e$ -— eccentricity,  $\dot M_\mathrm{tot}$ -- mass loss rate in the system, which in our case is connected only with the stellar wind from the parent star  
($\dot M_\mathrm{tot} \equiv  \dot M_\star$), 
$M_\mathrm{tot}$ -- total mass of the star~-~planet system (for our systems $M_\mathrm{tot}\approx M_\star$, where $M_\star$ -- is the stellar mass), $n$ -- mean motion ($n = 2\pi\sqrt{M_\mathrm{tot}/a^3}$). 

The system of equations is complemented with equations for the orbital inclination $i$, ascending node longitude $\Omega$, periapsis longitude $\varpi$, and periapsis argument $\omega$ \cite{greatescape}:


$$
\frac{di}{dt} = \frac{d\Omega}{dt} = 0,
$$

$$
\frac{d\varpi}{dt} = \frac{d\omega}{dt} = \frac{\sin(f)}{e} \frac{\dot M_\mathrm{tot}}{M_\mathrm{tot}}.
$$

The system of these equations does not have a complete analytical solution, but there are mass loss regimes under which an analytical solution is available. We are interested in one of these regimes. To classify them a dimensionless mass loss parameter $\psi$ is introduced. It is defined as follows:


\begin{equation}
    \psi = \frac{1}{2\pi} \left(\frac{a}{1 \mbox{au}}\right)^{3/2} \left(\frac{M_\star}{M_\odot}\right)^{-3/2} \frac{\dot M_\star}{M_\odot \, \mbox{yr}^{-1}} .
    \end{equation}

For $\psi\ll 1$ an adiabatic regime takes place. Then the evolution of the orbit is slow and can be described by a simple analytical formula:

\begin{equation}
\label{a_rude}
    a(\Delta t) = a_\mathrm{in} \left(1 - \frac{\Delta t \dot M_\star}{M_\mathrm{tot}}\right)^{-1}.
\end{equation}
Here $\Delta t$ is a duration of the evolutionary stage, $a(\Delta t)$ -- semimajor axis at the end of the evolutionary stage, $a_\mathrm{in}$ -- semimajor axis at the beginning of the evolutionary stage, $M_\mathrm{tot} $ -- current total mass of the star-planet system (the planet mass is considered to be constant).

The stellar mass evolution is calculated as follows:

\begin{equation}
\label{star_mass_ev}
    M_\star(\Delta t) = M_\mathrm{in} - {\Delta t \dot M_\star},
\end{equation}
where  $M_\star(\Delta t)$ is the stellar mass at the end of the evolutionary stage,  $M_\mathrm{in}$ --- stellar mass at the beginning of the evolutionary stage.

For cases where $\psi$ approaches unity (more precisely $\psi$>0.1) we numerically solve a system of four differential equations, three of which are given above, and the fourth describes the evolution of the mass loss rate (see section 3).


\subsection{Initial distributions of planetary parameters}

The key point in our modeling is the choice of initial distributions of planetary parameters. At the moment, they are not well known, so one can use different approaches to specify them.
 For example, as a basis for the distribution of exoplanets by masses and semimajor axes, one can use the data from one of the catalogs of confirmed exoplanets. However, modern observational data are burdened by various selection effects.
 Therefore, we decided to use the results of theoretical modeling of planetary systems.


In recent years, population models of the formation of planetary systems have been significantly advanced. In our simulation, when creating the initial distribution of planets on the plane ``semimajor axis of the planet's orbit~-~planetary mass'' ($a$~-~$M_\mathrm{pl}$), we rely on the results from the article by Alibert et al. \cite{alibert13}. 
In this paper, the authors calculated distributions over masses and semimajor axes at the end of the rapid initial dynamical evolution of planetary systems. The emphasis is on the fact that the calculations were carried out taking into account the interactions between planetary embryos and planets. The initial orbits of the embryos ranged from 0.1 to 20 AU, the initial masses were 0.01 Earth mass. The mass of the central star was taken equal to one solar mass. The metallicity of the star was chosen randomly from the metallicities of the stars in the CORALIE list of objects. The inner radius of the disk was taken to be 0.05 AU, the mass of the disks was in the range from 0.01 to 0.03 $M_\odot$, and the surface density at a distance of 5.2 AU -- from 0 to 10 $\mbox{g} / \mbox{cm}^2$ with a long ''tail'' of distribution up to 50 $\mbox{g} / \mbox{cm}^2$.


Following the approach used by Popkov and Popov \cite{PP2019} 
we approximate the $a$~--~$M_\mathrm{pl}$ diagram presented in \cite{alibert13} by several groups (see Fig.~\ref{fig:init_a_mp}).


\begin{figure}[t]
\centering
\includegraphics [scale=0.50]{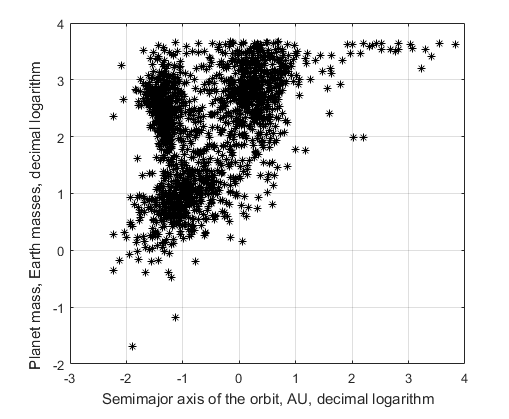}
\caption{Distribution of confirmed exoplanets in the plane ``semimajor axis of the planetary orbit~-~planetary mass'' according to the catalog exoplanet.eu. The figure shows data for approximately 1700 planets, including those for which only the lower limit $M\sin(i)$ is known as a mass. Planets around pulsars and white dwarfs are not included.}
\label{fig:eu_mp_a}
\end{figure}

Each of the groups I, IV -- VI is approximated by a two-dimensional log-normal distribution, which consists of two one-dimensional distributions:

\par
\begin{equation} p(x) = \frac{1}{x\sqrt{2\pi\sigma^2}}\exp\left(-\frac{(\ln(x) -\zeta)^2}{2\sigma^2}\right),
\end{equation} 
where $\sigma$ and $\zeta$ are parameters of the distributions, see table \ref{tab:init_distr}.

Group II is approximated by a bivariate Gauss distribution of the following form:

\begin{equation} 
\begin{split}
p(x,y) = \frac{1}{2\pi x y \sigma_x \sigma_y\sqrt{1-\rho^2}} \times 
\\
\times \exp\left(-\frac{\phi^2/\sigma_x^2+\psi^2/\sigma_y^2-2\rho\psi\phi/\sigma_x\sigma_y}{2(1-\rho^2)}\right),
\end{split}
\end{equation} 
where $\phi= \lg(x) - \zeta_x$, $\psi= \lg(y)- \zeta_y$, and $\sigma_x, \sigma_y, \zeta_x, \zeta_y, \rho$ are  parameters given in table \ref{tab:init_distr}.

Group III is approximated by a two-dimensional uniform (in a logarithmic scale) distribution.

Since the paper \cite{PP2019} focuses on those planets that can merge with their stars, i.e. on planets situated relatively close to their stars, the authors limit their modeling to the listed six groups of planets, which describe most of the distribution obtained by Alibert et al. in the $a$~---~$M_\mathrm{pl}$ plane. Analysis of Fig.~5 in \cite{alibert13} allows us to state that it has one more small group of planets (in our Fig. \ref{fig:init_a_mp}  this is group VIII) --- objects that finally appear at wide orbits as a result of gravitational interaction with other bodies in multiplanetary systems at early stages of their lives. According to \cite{alibert13}, the fraction of such planets in the population considered by us is slightly less than 1\%.


\begin{figure}[t]
\centering
\includegraphics [scale=0.45]{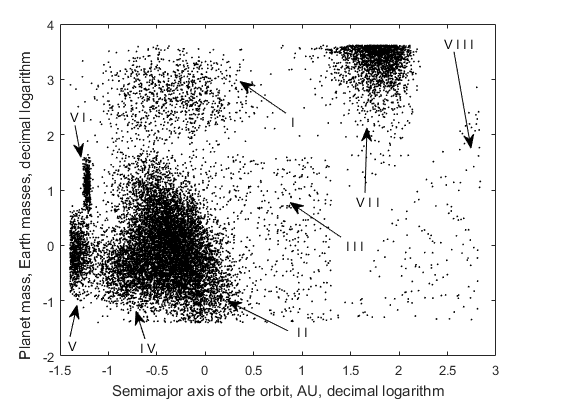}
\caption{Initial distribution of planets by masses and semimajor axes of the orbit (15,000 planets are shown).}
\label{fig:init_a_mp}
\end{figure}

This group is described by a two-dimensional ''triangular'' distribution uniform in a logarithmic scale, in which the ''hypotenuse'' is given by a straight line connecting the points with coordinates ($\lg (20)$, $\lg(0.1)$) and ($\lg(700)$, $\lg(1200)$), and ''legs'' are defined in the table \ref{tab:init_distr}.


\begin{figure}[t]
\centering
\includegraphics[scale=0.45]{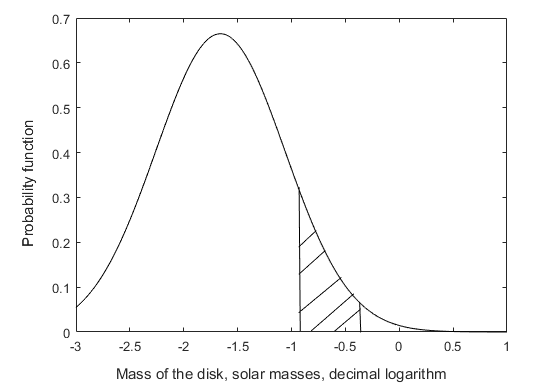}
\caption{Mass distribution of circumstellar disks according to \cite{mordasini09}. Distribution parameters: $\sigma=0.6, \mu = -1.66$. The mass interval in which planets can form in a self-gravitating disk is noted.}
\label{fig:discmasses}
\end{figure}

We also added a group that is not presented in \cite{alibert13}. These are the planets formed as a result of fragmentation of a self-gravitating protoplanetary disk. In describing this group, we follow the study by Forgan et al. \cite{forgan17}. Approximating the mass distribution of circumstellar disks by a normal distribution on a logarithmic scale, as was done in  \cite{mordasini09} (Fig. 4 and Table 2 in that paper and Fig. \ref{fig:discmasses} here) and assuming that the distribution is typical for the entire range of stellar masses in our work, we calculate the fraction of planetary systems with planets formed as a result of self-gravity of disk fragments. It should be noted that several assumptions are made in these calculations based on the results from \cite{forgan17}. Following this work, firstly, such planets can form only in disks with masses in the range from 0.125 to 0.4 stellar masses (lighter disks are unlikely to be self-gravitating, and heavier ones accrete onto the star very quickly). Secondly, the average number of disk fragments that can become planets in a given system was assumed to be equal to four. Finally, on average, about 40\% of these fragments survive. Taking into account these assumptions, the fraction of $\delta$ planets of this group in the population was calculated using the following equation:


\begin{equation}
\begin{split}
\delta = 0.4\times 4\times \\
\times \frac{\int_{\lg(0.125)}^{\lg(0.4)} \exp\left(-\frac{(x+1.66)^2}{2\times 0.6^2}\right)/\sqrt{2\pi \times 0.6^2} dx }{\int_{\lg(0.001)}^{\lg(1)} \exp\left(-\frac{(x+1.66)^2}{2\times 0.6^2}\right)/\sqrt{2\pi \times 0.6^2} dx}\\
\approx 0.1395.
\end{split}
\end{equation}

When choosing functions for the distributions of this group of planets we used data presented in Fig.3 and Fig.7 in \cite{forgan17}  as a guide. The distribution of semimajor axes  is approximated by a function similar to the Maxwell distribution:


$$
f(x)= \sqrt{\frac{2}{\pi}}\frac{x^2}{\kappa^3}\times\exp(-\frac{x^2}{2\kappa^2}).
$$
Parameter $\kappa$ is given in table \ref{tab:init_distr}. The mass distribution is taken to be log-normal.

The fraction of each of the first six groups of planets is calculated according to \cite{PP2019} (and corrected accordingly to account for the existence of two more groups). The percentage of all groups is given in the fourth column of table \ref{tab:init_distr}.


Regardless of which of the groups a planet belongs to, for the entire population, the lower and upper limits on the mass are determined. The lower limit is 0.04 Earth masses.  The upper one is  13 Jupiter masses (i.e., about 4120 Earth masses). The lower limit roughly corresponds to the mass of the least massive planet in the Solar system (we also note that at the moment only three exoplanets with masses below this limit are known, see e.g. exoplanet.eu). The upper limit is related to the lower limit on the mass of brown dwarfs.


For numerical calculations of planetary orbits for large values of the $\psi$ parameter, the values of eccentricity and true anomaly are required. For the entire population, we take the initial value of the true anomaly $f=0$, and the distribution of the initial eccentricity is set to be uniform in the range $0.01<e<0.1$. In our opinion, observational and/or theoretical data do not allow one to specify the distribution of this parameter with sufficient accuracy. Fixing the initial eccentricity at $e=0.1$ and $e=0.01$ led to the following change in the key results: the number of absorbed planets changed at the level of $\lesssim$ 1\%, the number of ejected planets -- at the level of $\lesssim$ 0.04\%.


\begin{table*}
\centering
\caption{Groups of initial distributions of planets in the plane $a$~-~$M_\mathrm{pl}$ and their parameters (units of measurement -- AU and Earth's masses, unless otherwise indicated).}
\begin{tabular}{||c| c| c| c||} 
 \hline
 Group & Distribution & Parameters & Fraction of \\ && &planets\\ 
 \hline\hline
  &  & {\footnotesize $\zeta_{a} = \ln(0.5)$} &  \\ I & 2D Log-normal & {\footnotesize $\zeta_{M} = \ln(500)$} & 6.72\%\\ & & {\footnotesize $\sigma_{a} = 0.9$ } &\\ & & {\footnotesize $\sigma_{M} = 1$ } & \\
 \hline
  &  & {\footnotesize $\zeta_{a} = \lg(0.5)$} &  \\  &  & {\footnotesize $\zeta_{M} = 0$ } &  \\ II & Bivariate Gauss & {\footnotesize $\sigma_{a} = 0.25$ } &46.78\% \\ & & {\footnotesize $\sigma_{M} = 0.45$ } & \\ & & {\footnotesize $\rho = -0.8$ } & \\
 \hline
  &  & {\footnotesize $log10(a_{min}) = -0.7$ } &  \\III & Uniform in logarithm & {\footnotesize $\log10(M_{min}) = -1.39$ } & 5.19\%\\  &   & {\footnotesize $\lg(a_{max}) = 1.3$ } &\\ & & {\footnotesize $log10(M_{max}) = 1.6$ } & \\
 \hline
  &  & {\footnotesize $\zeta_{a} = \ln(0.2)$} &  \\ IV & 2D Log-normal & {\footnotesize $\zeta_{M} = \ln(0.4)$} & 17.61\%\\ & & {\footnotesize $\sigma_{a} = 0.5$ } &\\ & & {\footnotesize $\sigma_{M} = 0.8$ } & \\
 \hline
 &  & {\footnotesize $\zeta_{a} = \ln(0.045)$} &  \\ V & 2D Log-normal & {\footnotesize $\zeta_{M} = \ln(0.7)$} & 6.04\%\\ & & {\footnotesize $\sigma_{a} = 0.2$ } &\\ & & {\footnotesize $\sigma_{M} = 0.8$ } & \\
 \hline
&  & {\footnotesize $\zeta_{a} = \ln(0.06)$} &  \\ VI & 2D Log-normal & {\footnotesize $\zeta_{M} = \ln(12)$} & 2.72\%\\ & & {\footnotesize $\sigma_{a} = 0.05$ } &\\ & & {\footnotesize $\sigma_{M} = 0.5$ } & \\  
 \hline 
  & Normal in logarithm (for masses)& {\footnotesize$\zeta_{M}$ = 23 $m_\mathrm{Jup}$} &  \\ VII &  & {\footnotesize $\sigma_{M}$ = 20 $m_\mathrm{Jup}$ } & 13.95\% \\
  & Maxwellian (for orbits) & {\footnotesize a = 40} &  \\
 \hline
  &  & {\footnotesize $\lg(a_{min}) = \lg(20) $ } & \\ VIII & Triangle uniform in logarithm &  {\footnotesize $\lg(M_{min}) = \lg(0.04) $ } & 1\% \\ & & {\footnotesize $\lg(M_{max}) =\lg(1200)$ } & \\ && {\footnotesize $\lg(a_{max}) = \lg(700)$ }& \\  [1ex] 
 \hline
\end{tabular}
\label{tab:init_distr}
\end{table*}

\subsection{Evolutionary tracks}
\label{model-ev tracks}
For calculations of the evolutionary tracks we use the MESA package (Modules for Experiments in Stellar Astrophysics, Release 10398) 
\cite{MESA}. Evolutionary tracks of stars with metallicity $ Z = 0.02$ are constructed for the following initial masses: from 1 to 2.6 solar masses with a step 0.1 $M_\odot$, then --- for masses 2.8 $M_\odot$, 3 $M_\odot $ and 3.25$M_\odot$, and finally, for larger masses with a step 0.25 $M_\odot$ up to 8 solar masses. Tracks for less massive stars were not used in the simulation, because the red giant stage in these tracks is reached in a time exceeding the age of the Galaxy (see subsection ''Masses of white dwarfs'' in the section devoted to the results of the work). 

For calculations of the mass loss rate at the stage of a red giant (Red Giant Branch --- RGB), we apply the Reimers formula \cite{blocker}:


\begin{equation} 
\label{reimers}
\frac{\dot M_\star}{M_\odot \, \mbox{year}^{-1}}= 4\times10^{-13}\times \eta_{R}\times \left(\frac{L}{L_\odot}\right) \left(\frac{R}{R_\odot}\right) \left(\frac{M_\star}{M_\odot}\right)^{-1}. 
\end{equation} 
Here $L$ is the current luminosity of the star, $R$ --- the current radius of a star, and $M$ --- the current mass of a star (all variable are taken in solar units). $\eta_{R}$ is a free parameter: for stars with initial masses up to 3 solar masses its values are set in the range 0.1--0.7 for both the RGB and the asymptotic giant branch (AGB); for more massive stars we use same range for the RGB, but for AGB we take $\eta_{R}$ in the range from 0.5 to 7.    

To calculate the mass loss rate at the AGB stage, we use the Blocker equation  \cite{blocker}: 

\begin{equation}
\begin{split}
\frac{\dot M_\star}{M_\odot \, \mbox{year}^{-1}}= 4\times 10^{-13}\times \eta_\mathrm{R} \left(\frac{L}{L_\odot}\right) \left(\frac{R}{R_\odot}\right) \left(\frac{M_\star}{M_\odot}\right)^{-1} \times  \\
\times 4.83\times 10^{-9}\times \left(\frac{L}{L_\odot}\right)^{2.7} \left(\frac{M_\star}{M_\odot}\right)^{-2.1}.
\end{split}
\label{eq:blocker}
\end{equation}

\begin{figure}[t]
\centering
\includegraphics [scale=0.55]{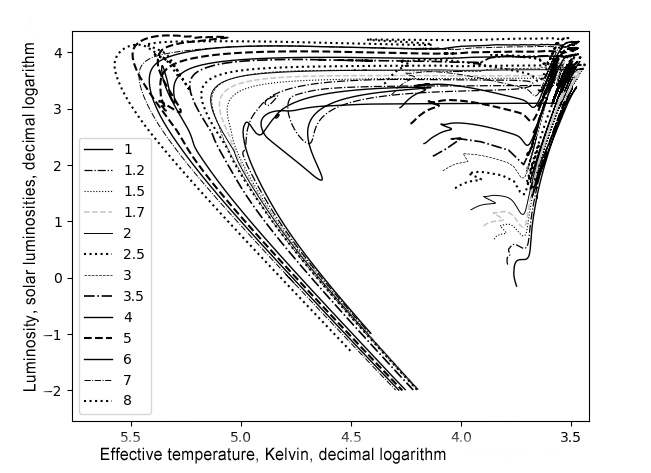}
\caption{Hertzsprung-Russell diagram. Evolutionary tracks from the MS stage to the white dwarf stage are shown. The initial masses are indicated.}
\label{fig:HR_full}
\end{figure}

 All the tracks used for modeling have been calculated up to the stage of white dwarf. The criterion for the onset of this stage is the drop in luminosity (due to cooling) below the critical value (see \ref{fig:HR_full}) after the end of the mass loss.
 
 Models of stars with an initial mass greater than 3 solar masses have also been brought to the stage of a white dwarf. In this case, the asymptotic giant branch in the models of massive stars used in this work ends with the stage of thermal bursts (Thermal Pulse Asymptotic Giant Branch --- TPAGB), which is limited to one or two short-term increases in luminosity (and mass loss rate). During them, a large part of the helium-hydrogen shell of the star is lost (see Fig. \ref{fig:6msol} and \ref{fig:6msol_lum} in the Appendix), after which the track ''turns'' towards an increase in the effective temperature, and the star passes to the planetary nebula stage where mass loss also occurs (several tenths of $M_\odot$ is lost then).
 

Evolutionary tracks for stars with an initial mass less than $3 M_\odot$ start at the pre-main sequence stage.  For more massive stars tracks start at the MS. Results of calculations of the stellar evolutionary track for each initial mass include a group of profile files, each of which describes the stellar structure at the corresponding moment of time, and a history file containing information about changes along the track in the main parameters of the star. The parameters whose changes are recorded in the history include the current age of the star, its effective temperature, its luminosity, mass, radius, mass loss rate, hydrogen and helium content in the center of the star, and many others. The history files of the tracks, which we calculated in MESA, contain information about changes in the main parameters of the star over numerous time intervals along the track. The number of these time points varies from $\sim 1200$ up to 30000 for different stellar masses. 
Most of these intervals correspond to the evolution of the star after the MS, including the asymptotic branch and the red giant branches.

From the parameters contained in the history files, for the population synthesis we directly need the mass loss rate, the radius of the star, and the corresponding age of the star, or the duration of the current stage of evolution. The main procedure of our population synthesis is a calculation of the planetary orbit at each evolutionary stage of the parent star. In this regard, one of the tasks is not to overload the program code with too much computation due to a large number of evolutionary stages. Another task is to monitor excluding some ``superfluous'' evolutionary stages, that the remaining stages and the corresponding mass loss rates give the same (within the error) final mass of the star (i.e., the white dwarf mass) which is obtained in the MESA calculation. The third task is not to exclude as a ``superfluous'' the stage when the radius of the star reaches its current maximum (because a growing star can swallow a nearby planet).

Solving these problems and working with the dependencies of the mass loss rate on time and of the stellar radius on time 
obtained from the data of history files, we make truncated versions of the tracks containing from 30 to 170 evolutionary stages, depending on the mass of the star. The longest ones are those where the largest number of thermal flashes occurs at the TPAGB stage (an example of a fragment of such track is shown in Fig. \ref{fig:2.3fragment} in the Appendix). The masses of white dwarfs obtained from the truncated tracks are systematically larger than the masses from the original tracks within 1\%. The first stage in each of the tracks turned out to be the longest. This is the MS stage. The rate of mass loss on the MS is calculated as the average value of the rate of loss at the beginning and at the end of the stage of MS. We consider the zero-age MS as the stage when the central content of hydrogen decreased by one hundredth compared to its initial content. The end of the MS is the moment when the central content of hydrogen becomes less than one-hundredth of the initial one. In all truncated tracks for each of the stages its duration, the average rate of mass loss at this stage, and the radius of the star at the end of the stage are prescribed. In some tracks of massive stars, there are stages whose duration does not exceed several years. This is done in cases when the rate of mass loss is very high (higher than $10^{-4} M_\odot$~{yr}$^{-1}$, see Fig. \ref{fig:6msol} in the Appendix).

\section{Population synthesis. The code}
\label{code}

The code is realized using the MatLab package. At the first step, we assign the total number of  ``star-planet'' pairs as one of the constants. This value determines the number of repetitions in the cycle of the procedures described below, as well as the number of planets of each of the groups in the $a$~--~$M_\mathrm{pl}$ plane. Each ``star-planet'' pair in the code randomly gets values of the semimajor axis of the planet's orbit, the mass of the planet, and the mass of the star. This is done by the pseudo-random number generator built into MatLab in accordance with the described initial distributions. The initial distribution of stellar masses is given by the Salpeter function:
$ dN/dM \sim M^{-2.3}$ \cite{Salpeter}, \cite{kroupa}). The generated masses of stars lie in the range from 1 to 8 $M_\odot$.
The initial eccentricity and the true anomaly of the planetary orbit are also defined. These values are chosen to be the same for all systems.
  

\begin{figure}[t]
\centering
\includegraphics [scale=0.5]{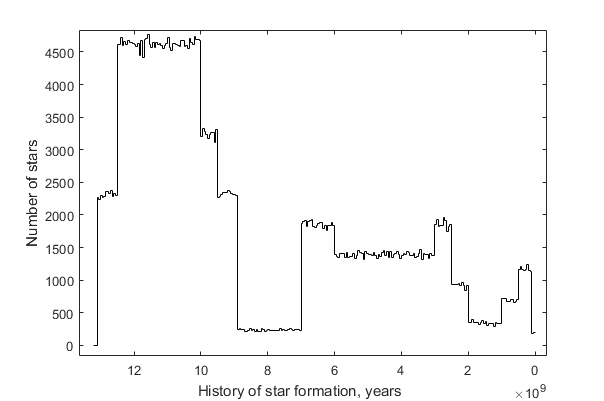}
\caption{The history of star formation in the Galaxy used in the simulation. The bin width is 50 million years. The number of stars in the sample is 500000. The normalization is made in such a way that the mass of all stars formed in the Galaxy in the range from 1 to 8 solar masses is equal to $19.55\times 10^{9} M_\odot$.} 
\label{fig:sfh}
\end{figure}

Also, at the first step, for each system, a bin in the star formation history (``age group'') is determined as well as the maximum possible age of the star corresponding to this group. To do this, the entire history of star formation in the Galaxy is divided into several stages (see Fig. \ref{fig:sfh}) with different rates of star formation, following the approximation in Fig. 1 in \cite{haywood}. It is assumed that throughout the history of star formation, the initial mass distribution of stars is given by the Salpeter function. For each stage, we calculate the ratio of the total mass of stars formed during this time interval to the total mass of stars in the Galaxy in the range from 1 to 8 solar masses. The latter is equal to $19.55\times 10^{9} M_\odot$, since the total initial mass of all stars in the Galaxy, formed during its lifetime, is assumed to be equal to $50\times 10^{9} M_\odot$, see \cite{haywood}. This ratio determines the range of random numbers corresponding to stars formed at a given stage in the history of star formation. Next, using a pseudo-random number generator, we get a value that determines the ''age group'' of the star. Then, again using the pseudo-random number generator, as well as conditional operators, we define the maximum possible age of the star.

Then the coefficient $N_{\rm{planets}}$ is calculated, which determines the number of planets around the star. The formula for calculating this coefficient is taken from \cite{PP2019}:



\begin{equation}
\label{planet_number} 
N_\mathrm{planets} = \begin{cases} 
 (M_{\star}/M_\odot)^{1.2} \times N_\mathrm{planet, sun} &\mbox{ if}\,  M_{\star} < 1.5 M_\odot; \\ 
 10, &\mbox{if}\,  M_{\star} > 1.5 M_\odot. 
 \end{cases}
\end{equation}
 Here $N_\mathrm{planet, sun} = 8$ --- the number of planets in the Solar system.
 This coefficient allows us to calculate the average number of planets around a star, it is used as an additional factor in obtaining the final distributions of planets (see eq.~\ref{planets_count}).

At the next step, the mass of the star in the current system is compared with the masses of the stars for which the evolutionary tracks are built, and for further work, the model with the closest mass is selected and the corresponding truncated track file is read. That is, the mass distribution is binned according to the selected values of the initial track masses calculated in MESA. The value of the semimajor axis of the orbit and the value of the mass of the star at the first evolutionary stage of the track are assigned equal to the initial values of the orbit and the mass of the star.

Next, the $\psi$ parameter is calculated. If its value is less than 0.1, then the value of the semimajor axis of the orbit and the mass of the star at the end of this evolutionary stage are determined by the formulas (\ref{a_rude}) and (\ref{star_mass_ev}), and the eccentricity and the true anomaly are not changed. If $\psi \geq$0.1, then the system of equations (\ref{dadt}) -- (\ref{dfdt}) with the condition $\dot M_\star=$~const is solved numerically with the Runge–Kutta method (RK4); the constant in this condition is determined by the value of the mass loss rate read from the file at this stage. The grid spacing is selected based on the duration of the evolutionary stage: the minimum spacing --- 0.1~yr, --- is chosen for very short stages and for stages when $\psi$ > 3. For stages longer than 100 years, a spacing of 5 years is used, with a stage duration of more than 1000 years --- 50 years, for all others --- 1 year.



After solving the specified system of equations, for the next evolutionary stage in addition to new values of the stellar mass and the semimajor axis of the planetary orbit, we define new values of eccentricity and true anomaly. Before moving on to the next evolutionary stage, the current value of the orbit is compared with the current radius of the star, and the current age, calculated as the sum of all past evolutionary stages, is compared with the maximum possible age.

If the value of the planet's pericentric distance and the current value of the stellar radius become equal or the stellar radius becomes larger than the planet's orbit, then the index of absorbed planets is increased by one.

The number of evolutionary stages in the track 
determines the number of repetitions of the calculation of the current mass of the star. It also determines the number of repetitions of the calculation of the value of the semimajor axis of the planet's orbit, if the value of $\psi$ does not exceed 0.1.

The calculations stop if the current age of the star at some stage reaches or exceeds the final age determined initially for it.





If the eccentricity reaches the value $e \geq$ 0.998 (in the middle or at the end of the evolutionary stage), then the planet is considered as escaped; the evolution of the elements of the orbit is suspended/stops (until the end of the system evolution on all at the next stages their values are preserved, fixed at the moment, when the eccentricity became greater than 0.998). The mass of the star continues to be calculated in accordance with the loss rates both at the current and at each of the following evolutionary stages using eq. (\ref{star_mass_ev}). For a test we varied the critical value of the eccentricity in the range from 0.99 to 0.999 -- this did not lead to a significant change in the number of runaway (and absorbed) planets. At the critical value of $e > 0.999$, we encounter instability in the code performance.

The values of the elements of the orbit, as well as the values of the mass and radius of the star, are saved at the end of each evolutionary stage (i.e., their values are also available after the completion of the entire program).

If the eccentricity value turned out to be less than zero at the current grid then the stage can be re-calculated with a reduced time step (down to 0.01 year). In such case, we numerically solve the system of equations (\ref{dadt}) -- (\ref{dfdt}).

For the final estimate of the Galactic population of escaped and absorbed planets and to obtain the final distributions of surviving planets in orbits and eccentricity (for the range of initial masses of stars stated above) we use the coefficient $N_\mathrm{planets}$ from eq. \ref{planet_number} and the coefficient $k$ (see eq. \ref{koef_k}). So the desired number $N$ corresponding to certain characteristics of the planets (for example, escaped or absorbed, etc.) in the Galaxy is determined as follows:

\begin{equation}
\label{planets_count}
 N = k\frac{N_\mathrm{calc} \sum_{n=1}^{N_\mathrm{tot}} N_\mathrm{planets, n}}{N_\mathrm{tot}}    
\end{equation}
where $N_\mathrm{calc}$ is the number of planets with the corresponding characteristics in the results of modeling, $N_\mathrm{tot}$ is the total number of ``star-planet'' pairs in the modeling (it is equal 500000 in all our runs). The coefficient $k$ is calculated as:

\begin{equation}
\label{koef_k} 
\begin{split}
k = \frac{\int_{1}^{8} M^{-2.3}dM}{N_\mathrm{tot}} \times \\ \times \frac{M_\mathrm{Gal}}{\int_{0.1}^{0.5} M^{-0.3}dM + \int_{0.5}^{150} M^{-1.3}dM},
\end{split}
\end{equation}
where $M_\mathrm{Gal}$ --- the total mass of all the stars in the Galaxy ($50\times 10^{9} M_\odot$, see \cite{haywood}), $N_\mathrm{tot} = 500000$, as it is in eq. \ref{planets_count}. For these parameters, we obtain the following value: $k\approx 17851$.



\section{Results}
\label{results}

\subsection{Orbit distribution. Escaped planets}

The modeling shows that about 60.2\% of the population of 500,000 planets turned out to be absorbed by parent stars at the red giant stage (RGB and AGB), 
about 0.3\%  
left their planetary systems and became free-floating planets. Using the eqs. \ref{planets_count} and \ref{koef_k} and basing on the obtained statistics of runaway planets, we estimate their number in the Galaxy. The obtained values are in the range of about 278~-~297 million. 


\begin{figure}[t]
\centering
\includegraphics [scale=0.45]{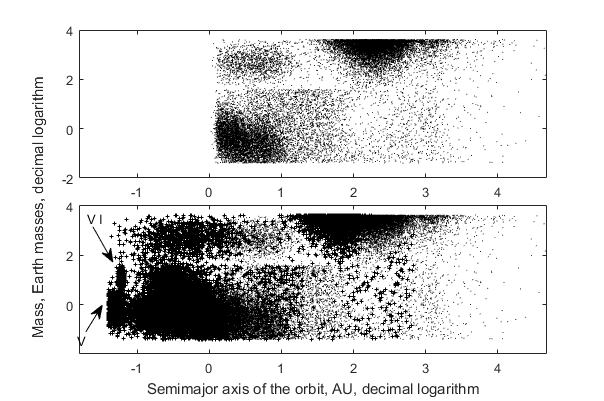}
\caption{Modeled distribution of planets by mass and semimajor axis at the final stage of evolution. Top: for the stars of the population that evolved to the white dwarf (WD) stage; bottom: for all stars. The ``+'' signs mark the planets around the stars that have not reached the WD stage. In order not to clutter up the figure, just 100000 points are shown, i.e. 20\% of the population considered in the modeling.}
\label{fig:a_mp}
\end{figure}

Moreover, in those ranges of stellar masses that are not covered in the code, according to our assumption, stars practically do not produce runaway planets --- either due to insufficient mass loss and stellar wind in the case of small stellar masses, or due to the short lifetime of the circumstellar disk in the case of massive stars with a powerful radiation flux \cite{6-8}.
 Of course, a certain number of planets can leave their systems due to dynamic interaction with other objects, but such a channel is not considered in our study.

\begin{figure}[t]
\centering
\includegraphics [scale=0.45]{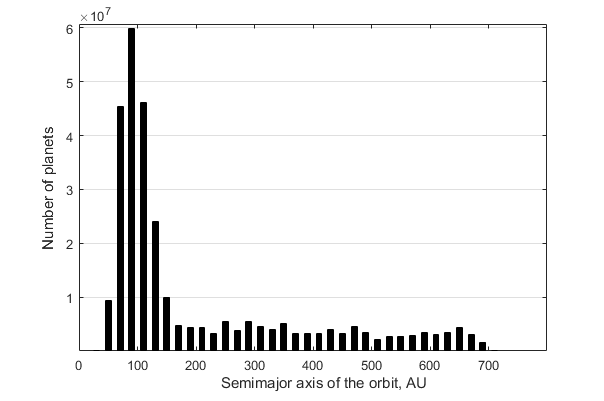}
\caption{Distribution of the initial orbits of runaway planets. The bin is 20 AU. The number of objects is normalised to the parameters of the Galaxy.}
\label{a_escaped}
\end{figure}

For the surviving planets around the stars that have gone through all the evolutionary stages and have become white dwarfs, the minimum values of the semimajor axis of orbit are observed for planets from the I-IV groups of $a$~--~$M_\mathrm{pl}$ distribution (the smallest value --- about 1.036 AU, --- for a planet from group II with initial semimajor axis of about 0.538 AU, and initial eccentricity $e \approx 0.01$, which has changed little during the life of the star). The maximal orbits are obtained for planets of groups VII and VIII; the runaway planets also appear only in groups VII and VIII. Most of the escaped planets have initial orbits close to 100 AU, see Fig. \ref{a_escaped}.



No planets of the groups V and VI (see also Fig. \ref{fig:init_a_mp}) survive around the stars that managed to evolve to the white dwarf stage, see Fig. \ref{fig:a_mp}. They are absorbed by the expanded envelopes of their parent stars at the giant stage. This is due to the fact that by the time they turn into a red giant, stars manage to lose such a fraction of their mass that the orbits of the planets increase little and the planets of the indicated groups situated close to the star are absorbed; the least massive stars shed a larger fraction of their mass on the RGB than on the AGB, but their radii also increase more significantly at this stage. It is also worth noting that among the population of stars that did not have time to evolve to the white dwarf stage, there are also those that swallowed up their planets.

\begin{figure}[t]
\centering
\includegraphics [scale=0.5]{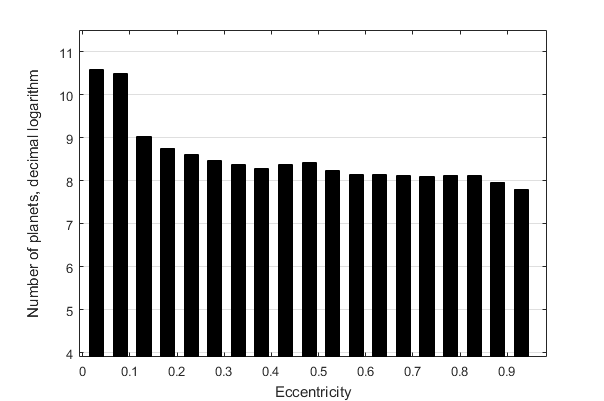}
\caption{Distribution of survived planets by the eccentricity for host stars that evolved to the white dwarf stage. The bin width is 0.05. The number of objects is normalized to the parameters of the Galaxy.}
\label{fig:ecc}
\end{figure}

A significant group of planets are moved to highly eccentric orbits (Fig. \ref{fig:ecc}). Since the formal criterion for a planet to leave its parent system is to reach $e = 0.998$, among the surviving planets there are several examples with an eccentricity $e > 0.99$ and an orbit of more than $10^5$ AU, and a couple of planets have orbits even more than a parsec. It is clear that such planets can be considered as bounded only according to the formal criteria indicated above, but taking into account, for example, Galactic tides they should be classified as “escaped”. However, in the statistics presented here, they are not included in the number of escapes. This is justified as the number of planets with orbits larger than \begin{math}10^4 \end{math} AU, but which did not formally leave their parent systems, turned out to be relatively small --  about 0.03$\pm 0.003$\% of the considered population, and in terms of the Galaxy --- about 30 million planets.

\begin{figure}[t]
\centering
\includegraphics [scale=0.5]{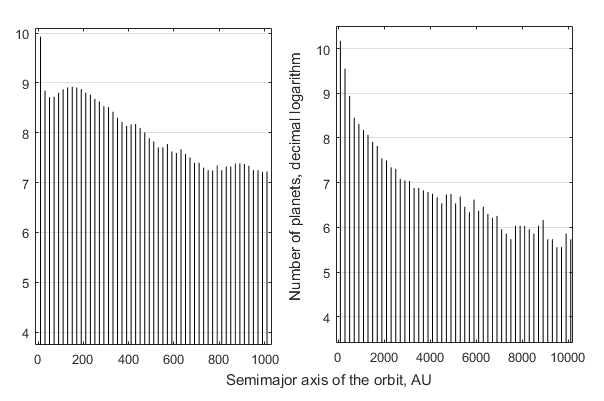}
\caption{Final distribution of planets by the semimajor axis of orbit for stars that have evolved to the white dwarf stage. Left: the bin width is 20 AU, Right: the bin width is 200 AU. The number of objects is normalized to the parameters of the Galaxy.}
\label{fig:afin_distr}
\end{figure}

The distribution of the remaining (survived) planets by orbits around white dwarfs is shown in Figure \ref{fig:afin_distr}. In particular, it shows the presence of a local maximum in the distribution of the number of planets in the region of 100-200 AU. This maximum is associated with the presence of a fairly large group of planets with large initial values of the semimajor axis of the orbit (see also Fig. \ref{fig:init_a_mp}, Fig.~\ref{fig:a_mp} and table \ref{tab:init_distr}).


\subsection{Future of the Earth}

Using the code described in this study we also perform modeling for the Earth-Sun system (the current age of the Sun is taken as 4.58 billion years).    

\begin{figure}[t]
\centering
\includegraphics[scale=0.5]{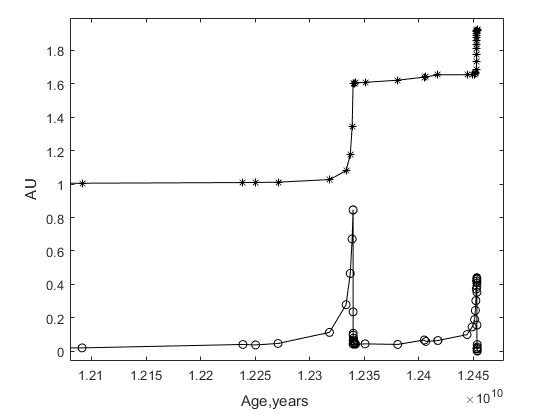}
\caption{The result of calculation of the Earth orbital evolution (the asterix symbols, upper curve) and the evolution of the solar radius (lower curve) using the MESA track for a star with the initial mass of 1$M_\odot$.}
\label{fig:earth}
\end{figure}

In our model, by the time the Sun turns into a white dwarf with a mass of about 0.52 $M_\odot$, the Earth will not be absorbed by the star at the red giant stage and will have a semimajor axis of about 1.922 AU. (Fig. \ref{fig:earth}). However, there are studies that show that the Earth will be absorbed by the Sun when the latter is at the red giant stage.
 Thus, according to calculations presented in the paper by Schroeder and Smith \cite{earth2008} the Earth will be swallowed up due to tidal effects, which are not taken into account in our work. However, judging by the maximum value of the radius of the Sun in our model, even accounting for the tides would not lead to the absorption of the Earth. The maximum value of the radius of a star with an initial mass of one solar mass that we obtained from the MESA tracks is inferior to the values given in the paper \cite{earth2008}: 255 $R_\odot $, or about 1.188 AU, versus 185 $R_\odot$, or 0.844 AU, in our model. It should be noted, however, that the evolutionary model for the Sun in the mentioned work was obtained for the metallicity $Z= 0.0188$, which is closer to the real solar metallicity than the value of $Z = 0.02$ used in all tracks in our simulation. We also note that in addition to the radii, the time that the Sun lives before it reaches the peak of the giant stage also differs (in the model by Schroeder and Smith this occurs approximately 20 million years earlier, compare Fig. \ref{fig:earth} in our work and Fig. 1 in \cite{earth2008}).


\subsection{Escaped planets and planets around massive stars and giants}

There are currently very few observational examples of planets around stars with masses of 3 or more solar masses in exoplanet databases. Confirmed examples include the following planets: o UMa b, Hip 79098 (AB) b, HD 17092 b, HD 13189 b, HD 119445, $\nu$ Oph b and c, BD+20 2457 b and c\footnote{exoplanet.eu/catalog} \footnote{\url{https://exoplanetarchive.ipac.caltech.edu/cgi-bin/TblView/nph-tblView?app=ExoTbls&config=PS}}. Moreover, the latter four are more likely to be brown dwarfs than planets. There are studies and observational programs devoted to the search for planets around giant stars, and no planet was discovered during one of such studies in which more than a hundred stars more massive than 2.7 $M_\odot$ were observed. Discussing these results, the authors suggest that the conditions in the protoplanetary disks around stars with initial masses above 2.5-3 $M_\odot$ are such that, in principle, giant planets cannot form there. At the same time, there are other studies devoted to the planets around stars with masses in the range from 6 to 8 solar masses \cite{6-8} and showing the theoretical possibility of the survival of planets around such stars.


\begin{figure}[t]
\centering
\includegraphics[scale=0.52]{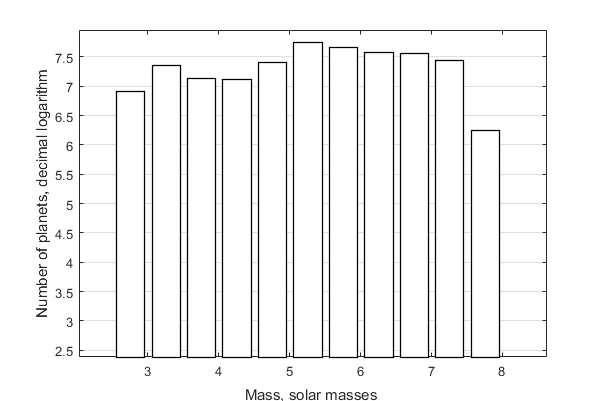}
\caption{Distribution of the initial masses of host stars whose planets were ejected from the parent systems. The bin width is 0.5 $M_\odot$. The number of objects is normalised to the parameters of the Galaxy.}
\label{fig:starmass_forescaped}
\end{figure}

Massive stars are of interest to us because they lose a sufficient amount of mass during their evolution prior to the white dwarf stage and achieve, at the corresponding stages of their evolution, such rates of mass loss that can lead to the loss of planets (see Fig. \ref{fig:starmass_forescaped}). In our simulation, the least massive star which ejected its planet due to mass loss by the end of evolution, has an initial mass of 2.6 $M_\odot$ (the initial orbit of this planet belonging to the VIII distribution group $a$~--~$M_\mathrm{pl}$, is $a_\mathrm{in} \approx 663$ AU). For the escaped planets from the VII distribution group $a$~--~$M_\mathrm{pl}$, the least massive stars were those with masses starting from 3.5 $M_\odot$.

\begin{figure}[t]
\centering
\includegraphics[scale=0.46]{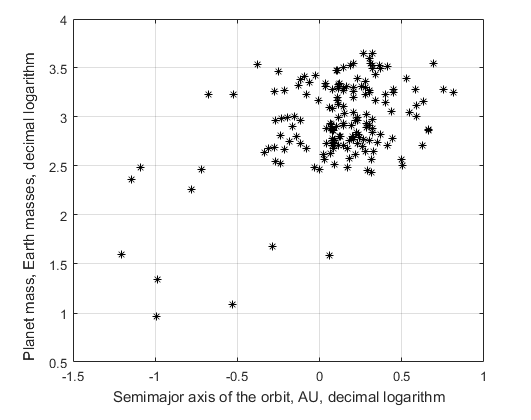}
\caption{Distribution of exoplanets in the $a$~--~$M_\mathrm{pl}$ plane around giants and subgiants according to the exoplanet.eu catalog.}
\label{fig:eu_giantstar_planets}
\end{figure}

Since the star is losing its mass, observational examples of interest may not be only stars more massive than 3 $M_\odot$, but also planets of slightly less massive stars that are at giant stage and have already lost some of their mass. There are many more examples of planets around giants than among stars more massive than 3 solar masses -- about 150 objects, discovered in most cases by variations in radial velocity. Most of these planets have orbits with semimajor axis less than 5 AU (Fig. \ref{fig:eu_giantstar_planets}). Examples of planets with a semimajor axis larger than 10 AU around giant stars have not been discovered, yet (although there are several examples among subgiants --- TYC 8998-760-1 b \footnote{\url{http://exoplanet.eu/catalog/tyc_8998-760-1_b}}, $ \kappa$ And b \footnote{\url{http://exoplanet.eu/catalog/kappa_and_b}}, 51 Eri b \footnote{\url{http://exoplanet.eu/catalog/51_eri_b}}).


Our simulation shows that on average, for escaped planets with small initial orbits the total mass ejected by the star should obviously be larger than for the planets initially remote from their stars. The closest of the escaped planets in our simulations has the initial orbit $a_\mathrm{in} \approx 52$ AU 
and eccentricity close to e = 0.1. It was ejected by the star with the initial mass of 7.5 $M_\odot$. 
 Thus, one can conclude that according to the observational data from exoplanet catalogs, a candidate for future escaped planets cannot be found among the already discovered planets. However, there are several examples of objects with high eccentricity among the confirmed planets of giant stars: HD 76920 b has an eccentricity $e =0.856$ and the semimajor axis $a = 1.15$ AU, HD 75458 b has $ e=0.713$ and $a = 1.275$ AU, HD 238914 b has $e = 0.56$ and $a = 5.7$ AU, HD 102272 c has $e = 0.68 $ and $a = 1.57$ AU,  HD 14067 b has $e = 0.533$and  $a = 3.4$ AU, Hip 97233 b has $e = 0.61$ and $a = 2.55$ AU,  HD 1690 b has $e = 0.64$ and $a = 1.3$ AU, Kepler-432 c has $e =0.64$ and $a = 1.188$ AU, BD+48 740 b has $e=0.76$ and $a = 1.7$ AU\footnote{http://exoplanet.eu/catalog}. Among the indicated systems there are those where the star has a mass of about 1.5 and even $\sim 2$~$M_\odot$. Depending on the rate at which these stars will lose most of this mass, the eccentricity of the planetary orbits in the future theoretically may turn out to be larger than unity, i.e. the planets will be no longer connected to their host stars. If we consider not only planets around evolved stars but also around MS stars, then we can also find candidates for future escaped planets.

\section{Discussion}
\label{disc}

\subsection{White dwarf masses}

\begin{figure}[t]
\centering
\includegraphics [scale=0.42]{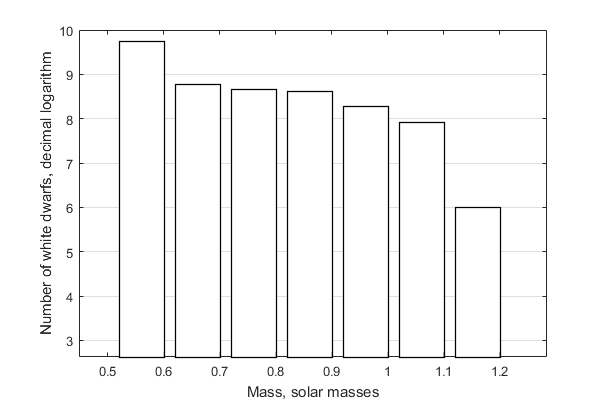}
\caption{Calculated white dwarf mass distribution for the stars with initial masses in the interval 1-8 $M_\odot$. The bin width is 0.1 $M_\odot$. The number of objects is normalized to the parameters of the Galaxy.}
\label{fig:wdlog}
\end{figure}

In the simulation, 82-83\% of the stars in the considered population manage to reach the white dwarf stage. The resulting mass distribution of white dwarfs is shown in Fig. \ref{fig:wdlog}. The most massive white dwarf obtained in our calculations has a mass of about 1.16 $M_\odot$, and the lightest -- 0.519 $M_\odot$. The comparison with modern theoretical and observational data on white dwarfs shows a relatively good correspondence between our results and the data if we take into account that our modeling does not include the evolution of low-mass stars of low metallicity (compare Fig. \ref{fig:wdlog} with the data from \cite {fieldwd}, \cite{sdssdr7} and \cite{wdmass}).

As it is mentioned in the subsection \ref{model-ev tracks} ``Evolutionary Tracks'', 
the lifetime of stars with masses below 1 $M_\odot$ and metallicity $Z=0.02$, before they become white dwarfs, exceeds the lifetime of the Galaxy, according to MESA calculations. Thus, for a star with an initial mass of 0.9 $M_\odot$, the lifetime on the MS is about 13.3 Gyr, and for such a star it takes about another 4.5 Gyr to reach the peak of the giant branch. A star with a mass of 0.95 $M_\odot$ lives on the MS for approximately 10.6 Gyr, and another 4 Gyr passes before it reaches its maximum radius on the giant branch.



We compare our results obtained in MESA with the results of calculations of evolutionary tracks carried out by different scientific groups. In particular, the article \cite{Charbonnel} presents evolutionary models of low-mass stars. In this study the following  MS lifetimes for stars with metallicity $Z=0.02$ and an initial mass of 0.8 and 0.9 solar masses are obtained: 22.7 and 14 Gyr, respectively. For a star with the initial mass 0.95 $M_\odot$, a comparison was made with the PARSEC tracks by the Padova group \cite{PARSEC}. After 13.7 -- 13.8 Gyr the central hydrogen abundance of a star of the indicated initial mass manages to fall by more than two orders of magnitude (which can indicate the end of the MS stage). But the star is still far from even reaching the stage of a giant, not to mention a white dwarf. Thus, comparison with modern advanced calculations of stellar evolution allows us to consider our choice of the mass interval of the white dwarfs' progenitors as appropriate.


\subsection{Evolutionary tracks}

Since the radii of stars and their mass loss rates at different stages of evolution obtained using the MESA package play a decisive role in the final statistics of the planets in our modeling, they deserve discussion and comparison with known calculations. We have already discussed the differences for stars with the mass 1 $M_\odot $ above.
For more massive stars comparison of our results with others is complicated due to the presence of the TPAGB stage where the radius increases repeatedly over  \textit{n} pulsations of the star, and not all the tracks are calculated to the end of TPAGB. Table \ref{rmax_compar} provides comparisons for some initial masses.


\begin{table}[t]
\caption{Comparison of the maximum stellar radii in different models}
\begin{tabular}{|l c c c|} 
 \hline  
  Initial mass & PARSEC & SSE & MESA \\ [0.5ex] 
\hline
 $2 M_\odot$ & 1.147 AU & 1.869 AU & 1.536 AU \\ [0.5ex]
 $5 M_\odot$ & 2.16 AU & 4.98 AU & 1.899 AU \\ [0.5ex]
 $6 M_\odot$ & 3.034 AU* & 5.97 AU & 2.252 AU \\ [1ex] 
 \hline
\end{tabular}
\\
* By the time TPAGB stage begins in a star
 with the mass $6 M_\odot$ the radius is 2.95 AU.
\label{rmax_compar}
\end{table}


Among the tracks constructed in MESA and used in the modeling, the maximum radius is achieved by a star with an initial mass of 6 $M_\odot $. It is $\approx$ 2.3 AU. Judging by our comparison, the maximum radii of the models obtained in MESA are smaller than in the SSE models. Many of the available PARSEC tracks do not reach the end of the TPAGB stage, thus comparison is not possible.

As for the mass loss rates, in MESA for the tracks of the most massive of the considered stars (from 3.5 $M_\odot $ up to 8 $M_\odot $) we obtained very large values at certain very short stages of evolution. This does not correspond to the existing observations. The largest value --- $10^{-2.2}$ $M_\odot$~{yr}$^{-1}$, ---
is obtained for models with initial masses 6.0 and $7.5 M_\odot$. In such cases, a star loses mass at this rate over a period of about fifty years. Note, that our models for stars with masses in the mentioned interval evolve from AGB to white dwarf stage losing their envelopes almost without thermal pulses (see Figs. \ref{fig:6msol_lum}, \ref{fig:6msol}), which is also not typical according to the observations. Still, in the case of massive stars, it is important for the fate of the planets that these stars lose much more than half of their mass during their evolution after the onset of the MS. Due to this, it becomes possible that the planets are no longer bound to their parent star.  

Calculations of stellar evolution with mass loss rates close to ours ($4\div7\times10^{-4}$  $M_\odot$~{yr}$^{-1}$) at the end of the AGB stage and during the ejection of a planetary nebula can be found in \cite{ mloss1983}, \cite{mloss1975}. The estimates of the stellar ``superwind'' for some of the observed OH/IR stars \cite{OH_Herschel}, \cite{OH_deJong}, \cite{OH2018}, \cite{OH_goldreich}  reach up to $\sim 10^{-3}$ $M_\odot$~{yr}$^{-1}$ . Note, that the famous Blöcker’s equation (see eq. (\ref{eq:blocker}) above) was proposed in the context of a discussion of high mass loss rates from Mira Ceti type stars and OH/IR stars. 

\subsection{Model development}

One way to improve the model is to consider the heterogeneity of the chemical composition of the stars in the population. As already indicated, all evolutionary tracks of stars in MESA are calculated for an initial metallicity of $Z = 0.02$. Focusing on modern models of the chemical evolution of the Milky Way, it is necessary to reflect the heterogeneity of the metallicity of the stellar population of the thin and thick disks and, possibly, the bulge of the Galaxy \cite{ChemEv}, \cite{3Im}. To do this, it is necessary to supplement the track grid by calculating the evolution of stars with metallicity $Z \approx 0.005$, corresponding to the peak of the distribution for thick disk stars (see Fig. 3 in \cite{3Im}) and, possibly, metallicity $Z \approx 0.04 $ for bulge stars (Fig. 5 in \cite{ChemEv}).

An important factor determining the statistics of planets obtained in the work is the assumptions made about their initial distribution in the $a$~--~$M_\mathrm{pl}$ plane. This distribution may be very different from the one used here. It is also important that the same distribution is used for different stellar masses. 
This is a significant simplification, which is made due to the lack of population calculation data differentiated by masses. Apparently, the corresponding data will appear in the nearest years (this is evidenced by the first works of a large cycle, that the Bern group \cite{PopSynth_new}, \cite{PopSynth3}, \cite{PopSynth2} began to publish).

Also, a very significant parameter for the statistics obtained in our study is the stellar mass loss rate. It should be noted that the decisive factor that prompted us to work with this grid of tracks calculated in MESA was the successful calculation of evolution up to the white dwarf stage and we obtained the final masses of stellar remnants directly. Thus, we did not have to resort to third-party sources or approximation formulas connecting the initial and final masses of stars in order to determine the mass of the star at the end of evolution for each track. If it is possible to obtain evolutionary tracks brought to the white dwarf stage with a more convincing evolution of mass loss rates, stellar radius, and other physical characteristics of stars then they will be used to improve the model.

In the current code, we do not take into account the influence of tides. We can try to estimate how the final statistics of absorbed and surviving planets will change, based on the obtained distributions and on data on the tidal absorption of planets by giant stars. However, calculations of tidal interaction still suffer from a number of uncertainties.

\begin{figure}[t]
\centering
\includegraphics [scale=0.45]{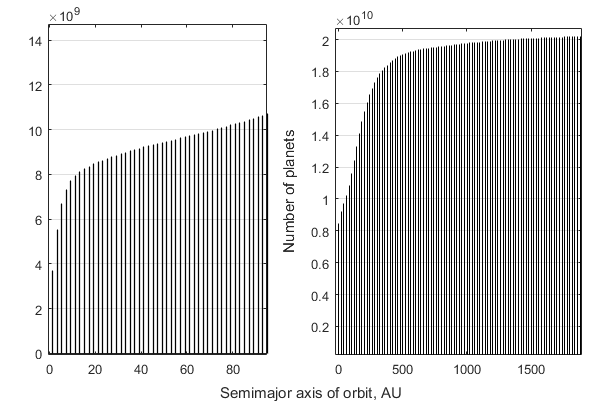}
\caption{Integral orbital distribution of the surviving planets around white dwarfs. Left: bin width is 2 AU; right: bin width is 20 AU. The number of objects is normalized to the parameters of the Galaxy.}
\label{fig:integr_afin}
\end{figure}

Tidal effects can affect the results of calculations for planets in close orbits around white dwarfs. 
In our simulation it turned out that at the end of evolution (at the white dwarf stage) there are a fairly large number of surviving planets in orbits close to their stars (Fig.\ref{fig:integr_afin}): 
within 2 AU --- about 3.7 billion 
which is less than one-fifth of all surviving planets (about 18\%), and within 4 AU --- about 30\%.
 However, the distribution of the masses of these planets shows that only a small fraction of the surviving planets close to their stars -- planets from group I -- have Jovian-scale masses. While among other surviving planets close to their parent stars, the average masses are about 1-5 Earth masses in group III (planets more massive than 0.15 Jupiter mass are not found in the results of calculations), and less than one Earth mass in group II (the most massive are also about 0.15 Jupiter mass). Such results give reason to believe that taking into account tidal effects \textit {with other assumptions unchanged} will increase the fraction of absorbed planets by no more than a few percent compared to our results since the role of tides is more important for massive planets.

Finally, there are a number of poorly known parameters associated with the general normalization of the number of planets. For example, we used a specific type of dependency presented in equation (12). Most likely in the future, for example, after increasing the statistics of known planets supplemented by the {\em Gaia} and {\em PLATO} satellite data, it will be possible to specify the number of planets in different types of systems with greater accuracy. For now, we have used a simplified form of the number of planets versus mass dependence, which leads to some systematic uncertainty in the total number of planets absorbed, survived, and ejected.

\section{Conclusion}
In this paper we presented population synthesis modeling of properties of planets at the late stages of stellar evolution.
Using the MESA package we model the evolution of stars from the Main Sequence stage to the formation of a white dwarf. We calculate the statistics of planets with different fates: absorbed, ejected from the system, and surviving by the time their parent stars transform into white dwarfs. We demonstrate that for the initial distributions of planets in the plane $a$~--~$M_\mathrm{pl}$ accepted in the work, the majority ($\sim$60\%) of planets born around stars in the mass range from 1~$ M_\odot$ up to 8~$ M_\odot$ is absorbed by their parent stars at the giant stage. A small fraction of planets ($\sim$0.3\%) is ejected from their systems due to the mass loss by their host star. We estimate the number of escaped planets with masses in the range from 0.04 Earth masses to 13 Jupiter masses in the Galaxy. It amounted to about 300 million objects.



\vskip 1cm
We thank the anonymous reviewer for useful comments that helped to improve  the manuscript.
The work is partially supported by the Interdisciplinary Scientific and Educational School of Lomonosov Moscow State University “Fundamental and Applied Space Research”. 

\vskip 1cm

{\it Translated by the authors}




\section*{Appendix}

This section provides examples of evolutionary tracks obtained in MESA and used in the study.


\begin{figure}
\centering
\includegraphics [scale=0.45]{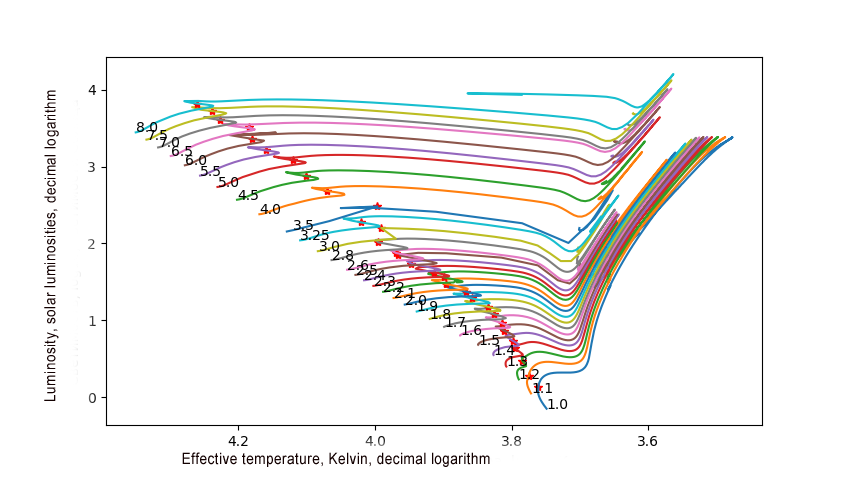}
\caption{Hertzsprung--Russell diagram for most of the tracks from the MS to the point of helium burnout in the center of the star.}
\label{helium_depletion}
\end{figure}

\begin{figure}
\centering
\includegraphics [scale=0.45]{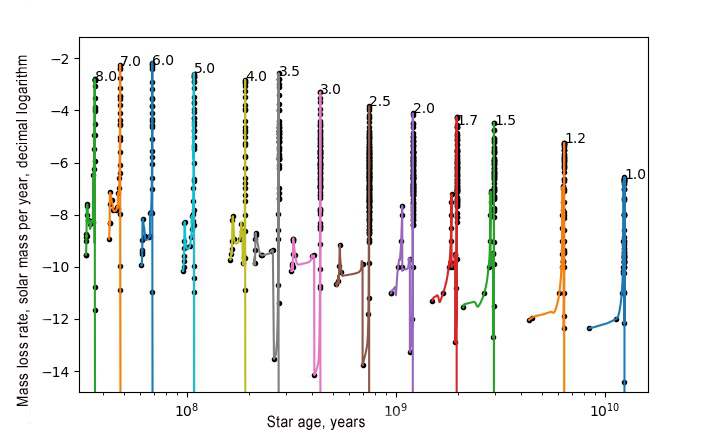}
\caption{Post-MS mass loss rates used in the modeling. The dots correspond to stages of the truncated tracks.}
\label{fig:massloss}
\end{figure}

Figure \ref{helium_depletion} shows the ''Effective temperature - Luminosity'' diagram of most of the tracks used in the simulation. The evolution from the Main Sequence stage to the stage of helium burnout at the center of the star is illustrated. 

Figure \ref{fig:massloss} shows the mass loss rates used in the modeling.

\begin{figure*}
\centering
\includegraphics [scale=0.7]{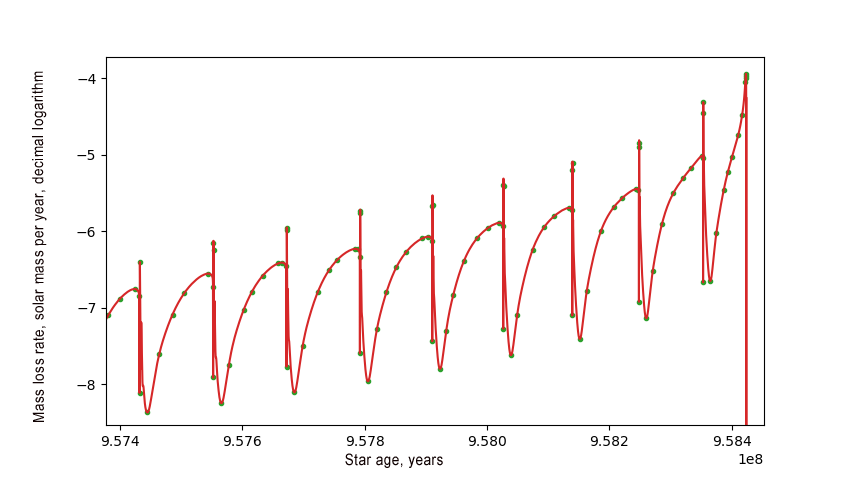}
\caption{A fragment of the track of a star with the initial mass 2.3 $M_\odot$ at TPAGB stage. The dots correspond to stages of the truncated tracks used in the modeling.}
\label{fig:2.3fragment}
\end{figure*}

Fig. \ref{fig:2.3fragment} illustrates the evolution of the mass loss rate for a star with an initial mass of 2.3 solar masses. The curve contains points corresponding to the truncated tracks described in the work: between each two neighboring points, there is a section corresponding to the evolutionary stage on which the mass loss rate is considered as the arithmetic mean of the loss rate values at these two points, i.e. at the beginning and at the end of the stage.




\begin{figure*}
\centering
\includegraphics [scale=0.7]{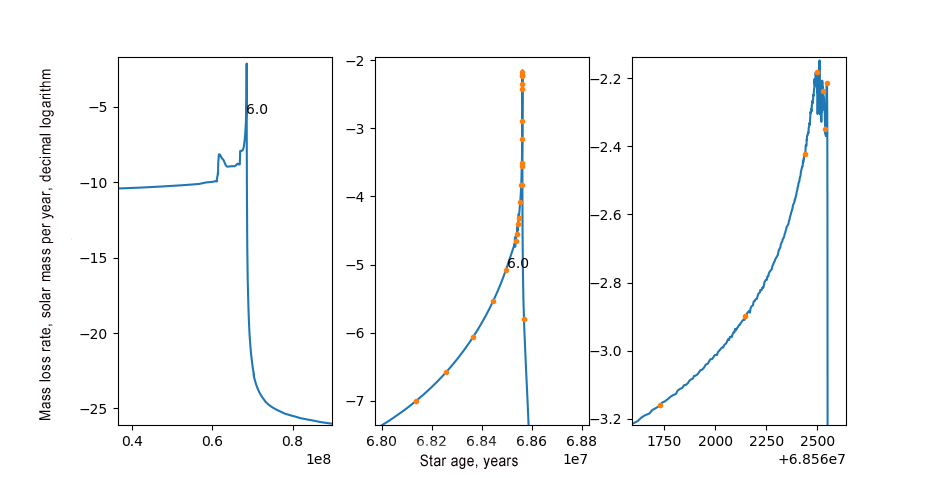}
\caption{Detailed evolution of the mass loss rate for a model with an initial mass 6$M_\odot$. In all three fragments the ordinate corresponds to the mass loss rate of the star and the abscissa corresponds to its age. The interval between each pair of points corresponds to one of the evolutionary stages of the truncated track.}
\label{fig:6msol}
\end{figure*}

%
In Fig. \ref{fig:6msol} we show in detail the evolution of mass loss rate for a star with an initial mass of 6$M_\odot$. Along with the model of a star with an initial mass of 7.5$M_\odot$, this model demonstrates the maximum local mass loss rate among all tracks. This rate is approximately equal to $10^{-2.2}$ $M_\odot$~{yr}$^{-1}$.


\begin{figure}
\centering
\includegraphics  [scale=0.7]{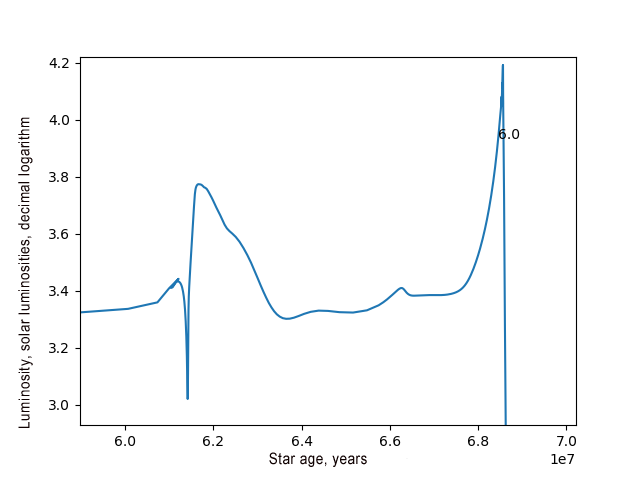}
\caption{Luminosity evolution in a model with an initial mass of 6$M_\odot$ for a part of the track. }
\label{fig:6msol_lum}
\end{figure}

Fig. \ref{fig:6msol_lum} shows the evolution of luminosity in the model with an initial mass of 6$M_\odot$. As indicated in the section \ref{model}, in MESA models of massive stars the TPAGB stage turned out to be represented by a single increase in luminosity, rather than a series of thermal flares.


\begin{figure}
\centering
\includegraphics [scale=0.7]{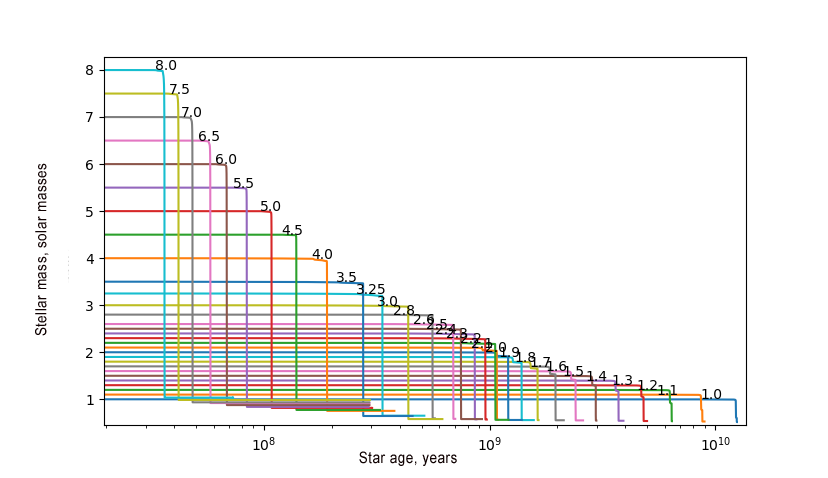}
\caption{Post-MS evolution of stellar masses for most of the tracks used in the study.}
\label{fig:stellar_mass}
\end{figure}

Figure \ref{fig:stellar_mass} illustrates the evolution of the masses in the tracks used in our simulation.


\begin{figure}
\centering
\includegraphics [scale=0.7]{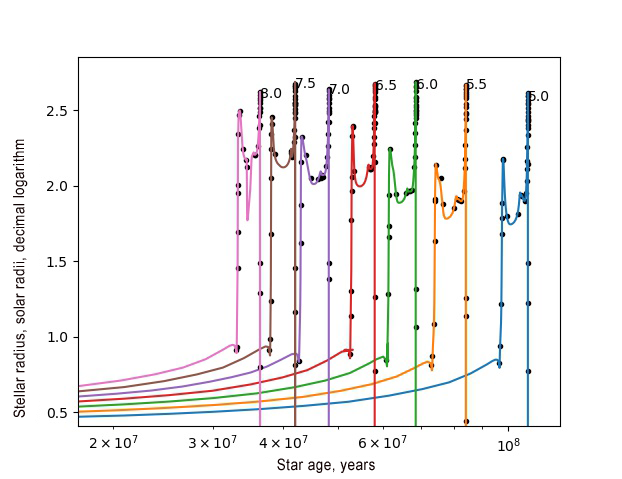}
\caption{Post-MS evolution of the stellar radii in massive stars models.}
\label{fig:lgR_massive}
\end{figure}

\begin{figure}
\centering
\includegraphics [scale=0.7]{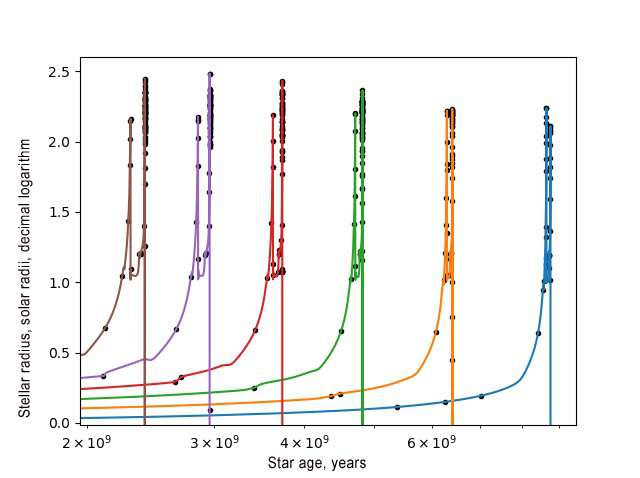}
\caption{Post-MS evolution of the stellar radii in models with initial masses 1.6, 1.5, 1.4, 1.3, 1.2 и 1.1 $M_\odot$ (right to left).}
\label{fig:lgR_fragment}
\end{figure}

Fig. \ref{fig:lgR_massive} and \ref{fig:lgR_fragment} show the evolution of the stellar radii for the models used in this study. The dots mark the values with which the semimajor axes of the planetary orbits are compared during the evolution of systems.


    

\end{document}